\documentclass[11pt]{article}
\usepackage[final]{graphics}
\font\tenimbf=cmmib10 at 10pt
\font\sevenimbf=cmmib10 at 6pt
\font\fiveimbf=cmmib10 at 4pt
\newfam\imbf
\textfont\imbf=\tenimbf
\scriptfont\imbf=\sevenimbf
\scriptscriptfont\imbf=\fiveimbf
\def\imb{\fam\imbf\tenimbf}
\begin{document}
%\draft
\begin{titlepage}
\title{\begin{center}
%%% {\Huge LAPTH}
\end{center}
%%% \vspace{5 mm}
%%% \hrule
%%% \vspace{20mm}
\bf{Photon  Production in Heavy Ion Collisions{\footnote {Talk given at the International Workshop on the physics of the quark-gluon plasma, 
Palaiseau, France, September 2001}}}}

\author{P.~Aurenche}
\maketitle

%%% \begin{center}
%%% \begin{enumerate}
\begin{center}
\item Laboratoire de Physique Th\'eorique LAPTH{\footnote {UMR 5108 du CNRS, 
associ\'ee \`a l'Universit\'e de Savoie}},\\ 
BP110, F-74941, Annecy le Vieux Cedex, France
\end{center}
%%% \end{enumerate}
%%% \end{center}

\vskip 3cm

\begin{abstract}

The production of  photons in heavy ion collisions is dicussed.

\end{abstract}
%%%   \vskip 4mm
\vfill
\centerline{\hfill LAPTH-Conf-893/2001}
\thispagestyle{empty}
\end{titlepage}

%%%% \maketitle
%
%\pagestyle{empty}
%

It has long been thought that the production of photons could be used to detect
the formation of the quark-gluon plasma  (QGP) in ultrarelativistic heavy ion
collisions. In fact, it is believed by some  that the present WA98 data
\cite{wa98} on inclusive photon production are evidence for QGP formation. At
CERN, the available center of mass energy of the nucleon-nucleon system is 
$\sqrt s \sim 17$ GeV. Since the rate of photon emission from the QGP phase is
expected to increase with energy, relatively to other production mechanisms, it
should become more and more relevant as one goes to RHIC ($\sqrt s \sim
.2$~TeV) and LHC ($\sqrt s \sim 5.5$~TeV).

The production of photons in heavy ion collisions is rather complex and one
roughly distinguishes four mechanisms: 

$-\ 1)$ the photon is produced in the hard interaction of two partons in the
incoming nuclei similarly to the well-known Chromodynamics (QCD) processes (QCD
Compton, an\-ni\-hi\-la\-tion, brems\-strahlung) in nucleon-nucleon collisions.
The rate is calculable in perturbative QCD and falls off at large transverse
momentum, $p_{_T}$, as a power law;

$-\ 2)$ in the collision of two nuclei the density of secondary hadrons is so
high that quarks and gluons become unconfined and a bubble of hot quark-gluon 
plasma is formed: at LHC the temperature is expected to be of the order of 1
GeV. Photons are emitted in the collisions of quarks and gluons in the plasma
with an energy spectrum which is exponentially damped but which should extend
up to several GeV;

$-\ 3)$  the QGP bubble expands and cools until a temperature of 150 to 200 MeV
is reached and a hadronic phase appears. As they collide the hot hadronic
resonances ($\pi^0,\ \rho,\ \omega$) emit photons until the freeze-out 
temperature is reached. The typical energy of such photons ranges from several
hundred MeV to several GeV;

$-\ 4)$ photons are decay products of resonances ($\pi^0,\ \eta,$ etc.)
emerging at the end of the thermal evolution (in which case their energy is of
the order of a few 100 MeV). $\pi^0$ and $\eta$ resonances can also be
produced at large $p_{_T}$ in hard parton scattering at the beginning of the
collision (their energy is of the order of a several GeV). These photons
together with those in 1) are a background to thermal photons produced in
2)-3).

A quantitative application of the above to the WA98 results leads to the
conclusion that it is ... impossible to conclude whether a QGP is formed!
Indeed the uncertainties on each mechanisms are such that the data can be
accomodated by  production of type 1) only (the ``German''
school~\cite{german}) or a mixture of type 1) and type 2-3) (the ``Indian'' 
school~\cite{indian} and the ``Finnish'' school \cite{finnish}). The problem is
that neither data nor theory tell us how to reliably  normalise, in the
$p_{_T}$ range of a few GeV, the rates from the various mechanisms.  In the
following we consider only inclusive photons of energy  in the GeV range where
thermal production (the ``signal") competes with the other mechanisms and we do
not deal with photon-hadron or photon-photon correlations which can also be
used to characterise the formation of QGP. We concentrate on the discussion of
uncertainties in the partonic production on the one end, and on the rate of
production in the quark-gluon plasma on the other hand. A review on
uncertainties associated to the hydrodynamic evolution of the hot matter can be
found in recent works, in particular~\cite{russian,finnish}.

\section{Partonic production of photons}

Since this mechanism is the same as that in proton-proton collisions, a lot of
what is known for this process should be relevant for heavy ions. The 
production cross section of direct photon in perturbative QCD is well known and
has the usual factorisable form \footnote {We do not discuss here the impact
parameter dependence of the collision.}: 
\begin{eqnarray}
%\begin{equation} 
{d \sigma^{^{AB}} \over d{\imb p_{_T}} dy} ={\sum}_{i,j,k} 
\int dx_a dx_b { dz \over z^2} F_{_{i/A}}(x_a,M) F_{_{j/B}}(x_b,M)  D_{\gamma/k}(z, M_{_F}) \nonumber \\
{d {\widehat \sigma}^{^{ij\rightarrow k}} \over d{\imb  p_k{_{_T}}} dy_k}   
(\mu, M, M_{_F}),
\end{eqnarray}
%\end{equation} 
where the functions $F$ are the parton densities and ${\widehat
\sigma}^{^{ij\rightarrow k}}$ is the hard cross section between partons $i$ and
$j$, in the nuclei $A$ and $B$, respectively to produce parton $k$. The function
$D_{\gamma/k}$ is the fragmentation function of parton $k$ into a photon. It
reduces to a $\delta(1-z)$ function in case the photon is produced directly ($k
= \gamma$) in the hard collision: in this case one refers to direct production
of the photon, otherwise one refers to bremsstrahlung production. The
calculations have been carried out up to next-to-leading order in QCD and they
suffer from the usual rather large ambiguities in the choice of the
renormalisation scale $\mu$, the factorisation scale $M$ and the fragmentation
scale $M_{_F}$~\cite{us-th}. Resumming ``threshold corrections" which is
appropriate for very large $p_{_T}$, reduces the dependence on
scales~\cite{threshold}. Extensive phenomenological studies have been carried
out for proton/(anti-)proton scattering for $\sqrt s$ from 20 GeV to 1.8 TeV
and the situation is very confused. Concerning data on inclusive photon
production it is impossible to accomodate all data within with the same set of
parameters: the E706 data (on Beryllium at 31.6 and 38.8 GeV) are at least a
factor 2 to 3 above the other data which extend from 20 to 63 GeV, and they do
not have the same $p_{_T}$ dependence~\cite{us-pheno}. There is therefore
little chance that present experiments can be used to normalise predictions for
future heavy ion collisions.

\vspace{.4cm}

Several further problems are specific to heavy ion collisions. They fall in two
categories. 

\subsection{Further theoretical uncertainties}

The kinematical domain of interest in heavy ion collisions is not exactly the
same as the domain where theory has been confronted to experiments. For WA98
data, both the energy and the relevant $p_{_T}$ range are rather low: is
perturbation theory reliable there? The present data in the same energy range
do not help either (E629 and NA3 \cite{na3}) as they have some relative
normalisation problems. For RHIC and LHC the interesting values of $p_{_T}$, to
distinguish mechanism 1) from 2-3) above, are a few GeV, {\em i.e.} very low
$p_{_T}$.
% at energies of .2 or 5.5 TeV. 
This introduces a ratio of large scales ($p^2_{_T}/ s$) which will
become very small and the standard NLO calculations are probably not reliable
then. This is the domain where ``recoil" resummation is important but further
studies are needed since the present results are rather dependent on
non-perturbative parameters \cite{recoil}. 

In the small $p_{_T}$ region, at high energies, the production of photons by
bremsstrahlung becomes overwhelming (for LHC, around 70\% at $p_{_T} \sim 5$
GeV/c decreasing to 40\% at $p_{_T} \sim 50$ GeV/c). However, the fragmentation
functions of a quark or a gluon into a photon \cite{bourhis} are not reliably
known in the relevant kinematical domain. Preliminary studies show a variation
of a factor 4 in the predicted cross section (at small $p_{_T}$) depending on
the two fragmentation sets used, both of which beeing equally good fits to
lower energy data \cite {guillet}.

\subsection{Nuclear effects}

Three effects can be considered: shadowing, ``primordial $k_{_T}$" and parton
energy loss. 

If shadowing is not a problem for fixed target experiments, such as WA98,
because they probe rather large $x$ values of the partons in the nuclei, it is
however predicted that it will reduce, at RHIC and LHC, the rate of production
of photons below $p_{_T} = 50$ GeV since the useful $x$ range (namely $x < .1$)
is well within the domain where shadowing occurs: at LHC the suppression can be
of the order of 30\% in the small $p_{_T}$ region~\cite{sarc}. Gluon initiated
processes play an important role, however not much is known about gluon
shadowing and to make predictions one has to rely on models~\cite{eskola} which
introduce further uncertainties in the predictions.

Nucleons in nuclei are expected to have some transverse momentum with respect
to the beam axis. Assuming that this transverse momentum is communicated to the
partons which scatter in type 1) processes, one has to modify eq. (1) to take
it into account. The effect is incorporated in leading logarithm calculations
in a simple way. In the usual implementation \cite{levai}, the main effect is
to modify the partonic cross section $\widehat \sigma$ which behaves as
$1/{\imb{p_{_T}}}^n$ in eq. (1) to $1/({\imb {p_{_T}-k^{^{(1)}}_{_T}-
k^{^{(2)}}_{_T}})}^n$ where the $\imb{k^{^{(i)}}_{_T}}$ are the intrinsic
transverse momenta of the incoming partons. Such a modification considerably
increases the size of $\widehat \sigma$ and consequently the photon production
rate specially if $\imb{p_{_T}}$ is comparable to $\imb{k^{^{(i)}}_{_T}}$. In
fact, a cut-off is introduced to prevent the vanishing of the denominator. To a
large extent, the primordial $k_{_T}$ effect is controled by the choice of this
arbitrary cut-off. Clearly, efforts should be made to improve the method and
provide a basis for reliable extrapolations. 

As mentioned above, at RHIC and LHC, bremsstrahlung production of photons
dominates the rate at low $p_{_T}$. But the final state parton looses energy 
\cite{bdms} in the hot medium ``before'' emitting the photon, thereby affecting
the energy spectrum of the photon and reducing its rate. In a recent model
calculation it was found that the reduction could be, very large, as much as
70\% for LHC  at $p_{_T} = 3$ GeV/c, becoming less drastic as $p_{_T}$
increases~\cite{sarc}. Further studies are necessary to include the latest
theoretical results and reduce the model dependence.

\section{Photon production in the quark-gluon plasma}

We assume the plasma in thermal equilibrium (temperature T) with vanishing
chemical potential. In principle, we could calculate the rate of production in
the plasma using  an expression similar to eq. (1), substituting to the nucleus
structure fonctions the Fermi-Dirac distribution for the quark and the
Bose-Einstein distribution for the gluon, not forgetting the appropriate
suppression or enhancement factors for partons in the final state. The matrix
elements for the various processes (QCD Compton, annihilation and
bremsstrahlung production) remain the same as before. A crucial difference with
eq. (1) is the phase space: there, the beam defines a longitudinal direction and
requiring the photon to be produced at large transverse momentum guarantees
that the internal lines in the hard matrix elements are finite (the
denominators in the matrix elements never vanish) and the production rate
remains finite. In the plasma, because quark and gluon distributions are
isotropic, there is no longitudinal direction with respect to which a minimum
transverse momentum for the photon can be imposed, and a naive use of eq. (1)
leads to a divergent result. The solution to this difficulty comes from the
fact that the partons in the plasma are not free but are in interaction with
other quarks and gluons in the medium: their properties are affected over long
distances or equivalently when they have soft momenta. The theoretical
framework to incorporate these effects is the Hard Thermal Loop (HTL) effective
theory \cite{braaten}.

Following the HTL approach one distinguishes two scales: the ``hard" scale,
typically of order $T$ or larger (the energy of quarks and gluons in the
plasma) and the ``soft" scale of order $g T$ where $g$, the strong coupling, is
assumed to be small. Collective effects in the plasma  modify the physics at
scale $g T$ {\em i.e.} over long distances of ${\cal O}(1/gT)$. These effects
lead to a modification of the propagators and vertices of the theory  and one
is led to introduce effective (re-summed) propagators and vertices. This is
easily illustrated with the example of the fermion propagator, $S(P)$, which in
the ``bare" theory is simply $1/p$ (we neglect spin complications and make only
a dimensional analysis). The thermal contribution to the one loop correction
$\Sigma(p)$ is found to be  $\Sigma(p)\sim g^2 T^2 /p$ which is of the same
order as the inverse propagator when $p$ is of order $gT$. The re-summed
propagator $^*S(P) = 1/ (p -\Sigma(p))$ is then deeply modified for momenta of
${\cal O} (gT)$. Likewise, the gluon propagator and vertices are modified by
hard thermal loops when the external momenta are soft. A practical consequence
of these thermal effects is to provide physical infrared cut-offs (effective
masses) to quarks and gluons which regularise the divergences mentioned above
in the naive approach. It also leads to new processes which increase the rate
of photon production~\cite{braatPY}. One can construct an effective
Lagrangian~\cite{braatP4} in terms of effective propagators and vertices and
calculate observables in perturbation theory.

 The rate of production, per unit time and volume, of a real
photon of momentum ($E,\imb p$) is calculated via the formula:
\begin{equation}
  {{E dN}\over{d{\imb p} dt d{\imb x}}}=-
  {1\over{(2\pi)^3}}\; n_{_{B}}(E)\,
  {\rm Im}\,\Pi^{^{R}}{}_\mu{}^\mu(E,{\imb p})\; ,
  \label{realphot}
\end{equation}
where $\Pi^{^{R}}{}_\mu{}^\mu(E,{\imb p})$ is the retarded photon
polarisation tensor. The pre-factor $n_{_{B}}(E)$ provides the expected
exponential damping $\exp(-E/T)$  when $E \gg T$. 
A loop expansion of $\Pi^{^{R}}$ is constructed with effective propagators and
vertices.

At one-loop, the photon production rate occurs via the usual QCD Compton and
annihilation processes the quarks having now an effective mass. The result is
finite and can be written as~\cite{baier}:
\begin{equation} 
{\rm Im}\,\Pi^{^{R}} (E,{\imb p}) \sim e^2 g^2 T^2
\Big(\ln ({E T\over m_q^2}) + C \Big) 
\label{eq:1loop}
\end{equation} 
where $m_q^2 \sim g^2 T^2$ is related to the thermal mass of the quark which
acts as the ``soft" cut-off . This leads to a ``large" logarithmic term of type
$\ln(1/g)$ dominating over a ``constant term" $C$.  The two-loop diagrams
contribute the processes shown in the figure.
\begin{figure}%3
\begin{center}
\centerline{\resizebox*{!}{3.75cm}{\includegraphics{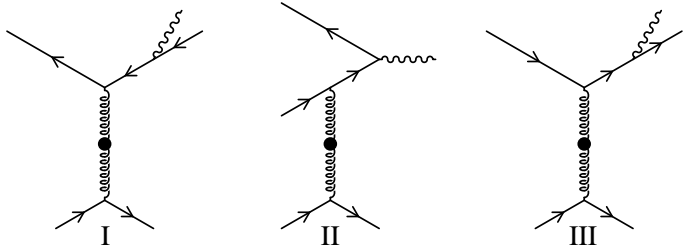}}}
\end{center}
\vskip -15pt
\caption{\em Physical processes occuring at two-loops.
      I:  bremsstrahlung from an antiquark. II: $q\bar{q}$ annihilation
      with scattering. III: bremsstrahlung from a quark.}
\label{fig:processes}
\end{figure}
 The result for hard photons is
\begin{eqnarray}
{\rm Im}\,\Pi^{^{R}} (E,{\imb p})\Big|_{\rm brems} \sim e^2\ g^2 T^2 \\
{\rm Im}\,\Pi^{^{R}} (E,{\imb p})\Big|_{\rm annil} \sim e^2\ g^2 T E
\label{eq:2loop}
\end{eqnarray}
These two-loop contributions have the same order as the one-loop
ones because of strong ``collinear" singularities~\cite{us-thermal-1} which
manisfest themselves as a factor of the form $T^2/m_q^2 \sim 1/g^2$: this factor
compensates the extra $g^2$ factor in the numerator. Another interesting result
of the calculation is the importance of process II in the figure: it describes
$q \bar q$ annihilation with scattering (a ``$3$ to $2$" process crossed from
bremsstrahlung) which grows with the energy of the photon and dominates over
the other contributions when $E/T \gg 1$.   Phenomenological applications of
these results have been carried out and the two-loop processes have been
included in hydrodynamic evolution codes to predict the rate of real photon
production. It is found that the two-loop processes (especially the
annihilation with scattering) lead to a large increase of the
rate~\cite{indian,finnish,russian,mustaT}. 

Since the one-loop and two-loop results are of the same order it is reasonable
to worry about the convergence of the perturbative expansion in the effective
theory. The enhancement mechanism operative at two-loop could also be at work
at the multi-loop level especially in ladder diagrams. It was indeed found that
the higher order ladder diagrams contribute to the same order as the two-loop
ones~\cite{us-thermal-2}.  Another effect which can modify the collinear
enhancement mechanism is related to the fermion damping rate: it can be
incorporated by  including a width on the fermion propagator which takes into
account the interaction length of the fermion in the plasma. This is
technically a higher order effect but it nevertheless modifies the rate of
production at the leading order~\cite{us-thermal-2}. It is interpreted as the
Landau-Pomeranchuck-Migdal effect~\cite{LandaP2}.  Both these higher order
effects have been combined in a consistent way~\cite{moore} and it is found a
reduction of the 2-loop photon production rate. From the phenomenological point
of view the reduction is less that 30\% in the $p_{_T}$ range of interest. All
these calculations are done for a plasma in chemical equilibrium which may not
be a realistic approximation: the effect of fugacities has been
considered in~\cite{mustaT,fugacity}: non linear effects occur since
decreasing the fugacities will reduce the rate of photon production but,
because the effective fermion mass is also reduced, the collinear enhancement
becomes stronger~\cite{fugacity}.

\vspace{.4cm}

In conclusion, the qualitative picture of the evolution of heavy ion collisions
seems rather well understood but a lot of progress has to be made for the model
to become really predictive. Only a few of the open problems have been
discussed here and there are many more such as the problems of thermalisation
and chemical equilibration which relate to the initial state conditions before
the hydrodynamic evolution takes place.

\vspace{.4cm}
\noindent
{\bf Acknowledgments}

I thank J.-Ph. Guillet and E. Pilon for discussions and F.~Gelis for
discussions and a critical reading of the manuscript. I also thank M. Gonin for the invitation to a very interesting and lively workshop.


\begin{thebibliography}{99}

\bibitem{wa98}
%OBSERVATION OF DIRECT PHOTONS IN CENTRAL 158-A/GEV PB-208 + PB-208 COLLISIONS.
WA98 Collaboration, M.M. Aggarwal et al.,
Phys.Rev.Lett.85:3595,2000; nucl-ex/0006007.

\bibitem{german} 
%A UNIQUE LARGE THERMAL SOURCE OF REAL AND VIRTUAL PHOTONS IN THE REACTIONS 
%PB-158-A/GEV + PB, AU. 
K. Gallmeister, B. Kampfer, O.P. Pavlenko, Phys.Rev.C62:057901,2000;\\  
%NUCLEAR BROADENING EFFECTS ON HARD PROMPT PHOTONS AT RELATIVISTIC ENERGIES. 
A. Dumitru, L. Frankfurt, L. Gerland, H. Stocker, M. Strikman,
Phys.Rev.C64:054909,2001;\\  
%HARD THERMAL PHOTON PRODUCTION IN RELATIVISTIC HEAVY ION COLLISIONS.  
F. D. Steffen, M. H. Thoma,  
Phys.Lett.B510:98,2001 
%BREMSSTRAHLUNG FROM AN EQUILIBRATING QUARK - GLUON PLASMA.
%By Munshi G. Mustafa (Giessen U.), Markus H. Thoma (CERN). Jan 2000. 27pp. 
%For erratum, see: HEPPH-0103293. 
%Published in Phys.Rev.C62:014902,2000  
%ERRATUM: BREMSSTRAHLUNG FROM AN EQUILIBRATING QUARK GLUON PLASMA.
%%By Munshi G. Mustafa, Markus H. Thoma. Apr 2001. 5pp. 
%Published in Phys.Rev.C63:069902,2001 

\bibitem{indian}
%PHOTONS FROM PB PB COLLISIONS AT CERN SPS.
Jan-e Alam, S. Sarkar, T. Hatsuda, T. K. Nayak, B. Sinha,
Phys.Rev.C63:021901,2001; \\ 
%SINGLE PHOTONS FROM PB+PB COLLISIONS AT CERN SPS, QGP VERSUS HADRONIC GAS.
A.K. Chaudhuri, 
nucl-th/0012058; \\
%RADIATION OF SINGLE PHOTONS FROM PB + PB COLLISIONS AT THE CERN SPS AND QUARK HADRON PHASE TRANSITION.
D. K. Srivastava, B. Sinha,
Phys.Rev.C64:034902,2001. 

\bibitem{finnish}
%PHOTON EMISSION IN HEAVY ION COLLISIONS AT THE CERN SPS.
P. Huovinen, P.V. Ruuskanen, S.S. R\"as\"anen,
nucl-th/0111052. 

\bibitem{russian}
%PHOTON EMISSION IN PB+PB COLLISIONS AT SPS AND LHC.
D.Yu. Peressounko and Yu.E. Pokrovsky, 
Nucl.Phys.A669:196,2000.

\bibitem{us-th}
%HIGHER ORDER QCD CORRECTIONS TO THE PHOTOPRODUCTION OF A DIRECT PHOTON AT HERA.
P. Aurenche, P. Chiappetta, M. Fontannaz, J.P. Guillet, E. Pilon,
Nucl.Phys.B399:34,1993;\\ 
%POLARIZED AND UNPOLARIZED PROMPT PHOTON PRODUCTION BEYOND THE LEADING ORDER.
L.E. Gordon, W. Vogelsang, 
Phys.Rev.D48:3136,1993. 

\bibitem{threshold}
%RESUMMATION OF THRESHOLD CORRECTIONS FOR SINGLE PARTICLE INCLUSIVE CROSS-SECTIONS.
E. Laenen, G. Oderda, G. Sterman, 
Phys.Lett.B438:173,1998;\\ 
%SUDAKOV RESUMMATION EFFECTS IN PROMPT PHOTON HADROPRODUCTION.
S. Catani, M. L. Mangano, P. Nason, C. Oleari, W.Vogelsang,
JHEP 9903:025,1999 

\bibitem{us-pheno}
%A CRITICAL PHENOMENOLOGICAL STUDY OF INCLUSIVE PHOTON PRODUCTION IN HADRONIC
% COLLISIONS.
P. Aurenche, M. Fontannaz, J.Ph. Guillet, B. Kniehl, E. Pilon, M. Werlen,
Eur.Phys.J.C9:107,1999; \\
%EXAMINATION OF DIRECT PHOTON AND PION PRODUCTION IN PROTON NUCLEON COLLISIONS
L. Apanasevich et al.,
Phys.Rev.D63:014009,2001. 

\bibitem{na3}
E629 Collaboration, 
M. McLaughin et al., 
Phys.Rev.Lett.31:971,1997;\\
NA3 Collaboration, J. Badier et al.,
Z.Phys.C31:341,1986.  

\bibitem{recoil}
%ORIGIN OF THE K(T) SMEARING IN DIRECT PHOTON PRODUCTION.
Hung-Liang Lai, Hsiang-nan Li,
Phys.Rev.D58:114020,1998; \\ 
%RECOIL AND THRESHOLD CORRECTIONS IN SHORT DISTANCE CROSS-SECTIONS.
E. Laenen, G. Sterman, W. Vogelsang,
Phys.Rev.Lett.84:4296,2000; Phys.Rev.D63:114018,2001. 

\bibitem{bourhis}
L. Bourhis, M. Fontannaz, J.-Ph. Guillet, 
Eur.Phys.J.C2:529,1998.

\bibitem{guillet}
J.-Ph. Guillet, private communication.

\bibitem{sarc}
%PROMPT PHOTONS AT RHIC.
J. Jalilian-Marian, K. Orginos, I. Sarcevic, 
Phys.Rev.C63:041901,2001; 
%NUCLEAR EFFECTS IN PROMPT PHOTON PRODUCTION AT THE LARGE HADRON COLLIDER.
%J. Jalilian-Marian, K. Orginos, I. Sarcevic, 
hep-ph/0101041.

\bibitem{eskola}
%THE SCALE DEPENDENT NUCLEAR EFFECTS IN PARTON DISTRIBUTIONS FOR PRACTICAL
%APPLICATIONS. 
K.J. Eskola, V.J. Kolhinen, C.A. Salgado,
Eur.Phys.J.C9:61-68,1999;\\
%NUCLEAR PARTON DISTRIBUTIONS IN THE DGLAP APPROACH.
K.J. Eskola, H. Honkanen, V.J. Kolhinen, P.V. Ruuskanen, C.A. Salgado,
Nuc.Phys.A661:645,1999.

\bibitem{levai} 
C.-Y. Wong, H. Wang, Phys.Rev.C58:376,1998;\\
%SATURATING CRONIN EFFECT IN ULTRARELATIVISTIC PROTON NUCLEUS COLLISIONS.
G.Papp, P. Levai, G. Fai,
Phys.Rev.C61:021902,2000; \\
%
%MULTIPLE SCATTERING AND P(T) BROADENING AT RHIC ENERGIES.
G. Papp, G.G. Barnafoldi, G. Fai, P. Levai, Y. Zhang,
%Talk presented at 15th International Conference on Ultrarelativistic
%Nucleus-Nucleus Collisions (QM2001), Stony Brook, New York, 15-20 Jan 2001.
nucl-th/0104021;\\
% 
%HIGH P(T) PION AND KAON PRODUCTION IN RELATIVISTIC NUCLEAR COLLISIONS. 
Yi Zhang, G. Fai, G. Papp, G. G. Barnafoldi, P. Levai,
hep-ph/0109233; 
%
%HARD PHOTONS AND NEUTRAL PIONS FROM RHIC.
%G.Papp, P. Levai, G. Fai,
%hep-ph/9904503. 

\bibitem{bdms}
%QUENCHING OF HADRON SPECTRA IN MEDIA
R. Baier, Y.L. Dokshitzer, A.H. Mueller, D. Schiff,
JHEP 0109:033,2001.

\bibitem{braaten}
E. Braaten, R. Pisarski, Nucl.Phys. B337 (1990) 569; B339 (1990) 310.

\bibitem{braatPY}
E. Braaten, R. Pisarski, T.C. Yuan, Phys.Rev.Lett.64,2242,1990

\bibitem{braatP4}
J.C. Taylor, S.M. Wong, Nucl.Phys.B346:115,1990;\\
E. Braaten, R. Pisarski, Phys.Rev.D45:1827,1992.

\bibitem{baier}
R. Baier, H. Nakkagawa, A. Niegawa, K. Redlich, 
Z.Phys.C53:433,1992. 

\bibitem{us-thermal-1}
%BREAKDOWN OF THE HARD THERMAL LOOP EXPANSION NEAR THE LIGHT CONE.
P. Aurenche, F. Gelis, R. Kobes, E. Petitgirard
Z.Phys.C75:315,1997; \\
%BREMSSTRAHLUNG AND PHOTON PRODUCTION IN THERMAL QCD.
P. Aurenche, F. Gelis, R. Kobes, H. Zaraket,
Phys.Rev.D58:085003,1998. \\
F. D. Steffen, M. H. Thoma, see \cite{german}. 

\bibitem{us-thermal-2}
%KLN THEOREM, MAGNETIC MASS, AND THERMAL PHOTON PRODUCTION.
P. Aurenche, F. Gelis, H. Zaraket,
Phys.Rev.D61:116001,2000; 
%LANDAU-POMERANCHUK-MIGDAL EFFECT IN THERMAL FIELD THEORY.
Phys.Rev.D62:096012,2000;\\
%A SIMPLE CRITERION FOR EFFECTS .....
F. Gelis, Phys.Lett.B493:282,2000.

\bibitem{LandaP2}
L.D. Landau, I.Ya. Pomeranchuk, Dokl.Akad.Nauk.SSR 92:735,1953\\
A.B. Migdal, Phys.Rev.103:1811,1956.

\bibitem{moore}
%PHOTON EMISSION FROM ULTRARELATIVISTIC PLASMAS.
P. Arnold, G. D. Moore, L. G. Yaffe,
JHEP 0111:057,2001;
%PHOTON EMISSION FROM QUARK GLUON PLASMA: COMPLETE LEADING ORDER RESULTS.
JHEP 0112:009,2001.

\bibitem{mustaT}
%BREMSSTRAHLUNG FROM AN EQUILIBRATING QUARK - GLUON PLASMA.
M. G. Mustafa, M. H. Thoma,
Phys.Rev.C62:014902,2000; Erratum, Phys.Rev.C63:069902,2001.

\bibitem{fugacity}
%HARD PHOTON PRODUCTION FROM UNSATURATED QUARK GLUON PLASMA AT TWO LOOP LEVEL.
D. Dutta, S.V.S. Sastry, A.K. Mohanty, K. Kumar, R.K. Choudhury, 
hep-ph/0104134 


%\bibitem{AltheR1}
%{T. Altherr, P.V. Ruuskanen}, {\it Nucl. Phys.} B {\bf 380} ({1992}) 377.
%{M.H. Thoma, C.T. Traxler}, {\it Phys. Rev.} D {\bf 56} ({1997}) 198. 

\end{thebibliography}
\end{document}